\documentclass[12pt]{article}
\usepackage{graphicx}
\usepackage{amsmath}
\usepackage{amssymb}
\usepackage{caption2}
\begin{document}
\bibliographystyle{prsty}
\begin{center}
{\large {\bf \sc{  Electromagnetic  form-factor
of the $\pi$ meson   with light-cone QCD sum rules  }}} \\[2mm]
Zhi-Gang Wang$^{1}$ \footnote{ E-mail,wangzgyiti@yahoo.com.cn.  }, Zhi-Bin  Wang$^{2}$     \\
$^{1}$ Department of Physics, North China Electric Power University, Baoding 071003, P. R. China \\
$^{2}$ College of Electrical Engineering, Yanshan University, Qinhuangdao 066004, P. R. China \\
\end{center}

\begin{abstract}
In this article, we calculate the electromagnetic form-factor of the
$\pi$ meson with the light-cone QCD sum rules.  The numerical value
$F_\pi^{p}(0) =0.999\pm 0.001$ is in excellent agreement with the
experimental data (extrapolated to the limit of zero momentum
transfer, or the normalization condition $F_\pi(0)=1$). For large
momentum transfers, the values from the two sum rules are all
comparable with the experimental data and theoretical estimations.
\end{abstract}


\section{Introduction}

The $\pi$ meson, as both Nambu-Goldstone boson and quark-antiquark
bound state, plays an important role in testing the quark models and
exploring the  low energy QCD. Its electromagnetic form-factor and
electromagnetic radius are important parameters, and have been
extensively studied  both experimentally
\cite{exp1,exp2,exp3,exp4,exp5,exp6,exp8} and theoretically, for
examples, the QCD sum rules
\cite{Radyushkin821,Radyushkin822,Radyushkin823,Radyushkin824}, the
light-cone QCD sum rules \cite{Braun94,Braun00,Bijnens02,Agaev05},
perturbative QCD
\cite{Radyushkin2004,Brodsky80,HnLi921,HnLi922,Huang04,Bakulev04},
Schwinger-Dyson equation \cite{SDE1,SDE2,SDE3}, etc\footnote{In
Ref.\cite{Radyushkin2004}, Radyushkin  introduces the distribution
amplitude of the $\pi$ meson for the first time, expresses   the
 form-factor of the $\pi$ meson in terms of the  distribution
 amplitudes asymptotically, and formulates the perturbative QCD parton picture
for hard exclusive processes. }.

In Refs.\cite{Braun94,Braun00,Bijnens02,Agaev05}, the axial-current
is used to interpolate the $\pi$ meson, in
Refs.\cite{Braun00,Bijnens02}, the  radiative $\mathcal
{O}(\alpha_s)$ corrections and higher-twist effects are taken into
account. In this article, we choose the pseudoscalar current to
interpolate the $\pi$ meson and calculate the electromagnetic
form-factor of the $\pi$ meson with the light-cone QCD sum rules. In
our previous works, we have studied the vector form-factors and
scalar form-factors of the $\pi$ and $K$ mesons, the form-factors of
the nucleons, and obtain satisfactory results \cite{Wang061,
Wang062, Wang063,Wang071,Wang072}. The light-cone QCD sum rules
carry out the operator product expansion near the light-cone
$x^2\approx 0$ instead of short distance $x\approx 0$, while the
non-perturbative matrix elements are parameterized by the light-cone
distribution amplitudes (which classified according to their twists)
instead of
 the vacuum condensates \cite{LCSR1,LCSR2,LCSR3,LCSR4,LCSR5,LCSRreview1,LCSRreview2}. The non-perturbative
 parameters in the light-cone distribution amplitudes are calculated
 with  the conventional QCD  sum rules
 and the  values are universal \cite{SVZ79,Reinders85}.

The article is arranged as: in Section 2, we derive the
electromagnetic  form-factor  $F_\pi(Q^2)$ with the light-cone QCD
sum rules; in Section 3, the numerical result and discussion; and in
Section 4 is reserved for  conclusion.

\section{Electromagnetic  form-factor  of the $\pi$ meson   with light-cone QCD sum rules}

In the following, we write down the definition  for the
electromagnetic form-factor  $F_\pi(q^2)$,
\begin{eqnarray}
\langle \pi(p_2)|J_\mu(0)|\pi(p_1)\rangle=F_\pi(q^2)(p_1+p_2)_\mu \,
,
\end{eqnarray}
where the $J_\mu(x)$ is the electromagnetic current and $q=p_2-p_1$.
We study the electromagnetic form-factor $F_\pi(q^2)$ with the
 two-point correlation function $\Pi_{\mu}(p,q)$,
\begin{eqnarray}
\Pi_{\mu}(p,q)&=&i \int d^4x \, e^{-i q \cdot x} \,
\langle 0 |T\left\{J_\pi(0) J_{\mu}(x)\right\}|\pi(p)\rangle \, ,\nonumber \\
J_\mu(x)&=&e_u{\bar u}(x)\gamma_\mu  u(x)+e_d{\bar d}(x)\gamma_\mu  d(x)\, ,\nonumber \\
 J_{\pi}(0)&=& \bar{d}(0)i\gamma_5u(0) \, ,
\end{eqnarray}
 where we choose the pseudoscalar current $J_\pi(0)$ to interpolate    the $\pi$ meson.
 The correlation function
$\Pi_{\mu}(p,q)$ can be decomposed as
\begin{eqnarray}
\Pi_{\mu}(p,q)&=&\Pi_{p}\left(p,q\right)p_{\mu}+\Pi_{q}
\left(p,q\right)q_{\mu}
\end{eqnarray}
due to  Lorentz covariance.  In this article, we derive the sum
rules with the tensor structures $p_\mu$ and $q_\mu$, respectively.

According to the basic assumption of the current-hadron duality in
the QCD sum rules approach \cite{SVZ79,Reinders85}, we can insert  a
complete series of intermediate states with the same quantum numbers
as the current operator  $J_\pi(0)$  into the correlation function
$\Pi_{\mu}(p,q) $ to obtain the hadronic representation. After
isolating the ground state  contribution from the pole term of the
$\pi$ meson,  the correlation function $ \Pi_\mu(p,q)$
  can be expressed in the following form,
\begin{eqnarray}
\Pi_{\mu}(p,q)&=&\frac{2f_\pi
m_\pi^2F_\pi^{p}(q^2)}{(m_u+m_d)\left[m_\pi^2-(q+p)^2\right]}p_\mu
+\nonumber\\
&&\frac{f_\pi m_\pi^2
F_\pi^{q}(q^2)}{(m_u+m_d)\left[m_\pi^2-(q+p)^2\right]}q_\mu+\cdots
\, ,
\end{eqnarray}
where we introduce the indexes  $p$ and $q$ to denote the
electromagnetic form-factor from the tensor structures   $p_\mu$ and
$q_\mu$ respectively, and we use the standard definition  for the
decay constant
   $f_{\pi}$,
\begin{eqnarray}
\langle0|J_{\pi}(0)|\pi(p)\rangle&=&
\frac{f_{\pi}m_{\pi}^2}{m_u+m_d}\, .\nonumber
\end{eqnarray}

 In the following, we briefly outline
the operator product expansion for the correlation function
$\Pi_\mu(p,q)$ in perturbative QCD theory. The calculations are
performed at the large space-like momentum regions $P^2=-(q+p)^2\gg
0$ and $Q^2=-q^2\gg 0$, which correspond to the small light-cone
distance $x^2\approx 0$   required by validity of the operator
product expansion approach\footnote{In the frame where the $\pi$
meson has a finite 3-vector $|\overrightarrow{p}|\sim \mu$, $\mu^2
\ll Q^2$, the $p_\mu$ and $q_\mu$ can be approximated as
$p_\mu=(\sqrt{m_\pi^2+\mu^2},0,0,\mu)\approx(\mu,0,0,\mu)$ and
$q_\mu=\left(\frac{\xi Q^2}{4\mu},0,0,\sqrt{(\frac{\xi
Q^2}{4\mu})^2+Q^2}\right)\approx \left(\frac{\xi
Q^2}{4\mu},0,0,\frac{\xi Q^2}{4\mu}+\frac{2\mu}{\xi}\right) $, where
$\xi\sim 1$, we obtain the relation $q^2\ll 0$ and $(p+q)^2\ll 0$.
$q \cdot x=q_0 x_0 -q_3 x_3\approx \frac{\xi Q^2}{4\mu}
(x_0-x_3)-\frac{2\mu}{\xi} x_3 $, we take the values $x_0-x_3 \sim
\frac{4\mu}{\xi Q^2}$ and $x_3 \sim \frac{\xi}{2\mu} $ to avoid
strong oscillation, $x^2\sim \frac{1}{Q^2}\rightarrow 0$. For more
details, one can consult Ref.\cite{LCSRreview2} }. We write down the
propagator of a massive quark in the external gluon field in the
Fock-Schwinger gauge firstly \cite{Belyaev94},
\begin{eqnarray}
&&\langle 0 | T \{q_i(x_1)\, \bar{q}_j(x_2)\}| 0 \rangle =
 i \int\frac{d^4k}{(2\pi)^4}e^{-ik(x_1-x_2)}\nonumber\\
 &&\left\{
\frac{\not\!k +m}{k^2-m^2} \delta_{ij} -\int\limits_0^1 dv\, g_s \,
\frac{\lambda^a_{ij}}{2} G^a_{\mu\nu}(vx_1+(1-v)x_2)
 \right. \nonumber \\
&&\left. \Big[ \frac12 \frac {\not\!k
+m}{(k^2-m^2)^2}\sigma^{\mu\nu} - \frac1{k^2-m^2}v(x_1-x_2)^\mu
\gamma^\nu \Big]\right\}\, ,
\end{eqnarray}
where the $G^a_{\mu \nu }$ is the gluonic field strength.
Substituting the above $u$ and $d$ quark propagators and the
corresponding $\pi$ meson light-cone distribution amplitudes into
the correlation function $\Pi_\mu(p,q)$, and completing the
integrals over the variables $x$ and $k$, finally we obtain the
representation at the level of quark-gluon degrees of freedom,
\begin{eqnarray}
\Pi_i(p,q)&=&e_u\Pi_i^u(p,q)+e_d\Pi_i^d(p,q) \, ,
\end{eqnarray}
the explicit expressions of the $\Pi_i^u(p,q)$ and $\Pi_i^d(p,q)$
are given in the appendix.  In calculation, we have used the
two-particle and three-particle   light-cone distribution amplitudes
of the $\pi$ meson
\cite{LCSR1,LCSR2,LCSR3,LCSR4,LCSR5,Belyaev94,PSLC1,PSLC2,PSLC3,PSLC4},
the explicit expressions of the light-cone distribution amplitudes
are also
  presented in the appendix. The parameters in the
light-cone distribution amplitudes are scale dependent and estimated
with the QCD sum rules
\cite{LCSR1,LCSR2,LCSR3,LCSR4,LCSR5,Belyaev94,PSLC1,PSLC2,PSLC3,PSLC4}.
In this article, the energy scale $\mu$ is chosen to be
$\mu=1\,\,\rm{GeV}$.

We take the  Borel transformation with respect to  the variable
$P^2=-(q+p)^2$   for the correlation functions $\Pi_p(p,q)$ and
$\Pi_q(p,q)$. After matching  with the hadronic representation below
the threshold, we obtain the following two sum rules for the
electromagnetic form-factors $F_\pi^{p}(q^2)$ and $F_\pi^{q}(q^2)$
respectively,

\begin{eqnarray}
&&\frac{2f_\pi m_\pi^2}{m_u+m_d}F_\pi^p(q^2)e^{-\frac{m_\pi^2}{M^2}} \nonumber\\
&=&  \frac{f_\pi m_\pi^2}{m_u+m_d}\int_{\Delta}^1du
\varphi_p(u)e^{-\Xi}-\frac{(e_um_u-e_dm_d)f_\pi
m_\pi^2}{M^2}\int_{\Delta}^1du \int_0^u
dt \frac{B(t)}{u}e^{-\Xi}  \nonumber\\
&&+\frac{1}{6}\frac{f_\pi m_\pi^2}{m_u+m_d}\int_{\Delta}^1du
\varphi_\sigma(u)\left\{\left[1-u\frac{d}{du}
\right]\frac{1}{u}+\frac{2(e_um_u^2-e_dm_d^2)}{u^2M^2}\right\}e^{-\Xi}
  \nonumber\\
  &&+(e_um_u-e_dm_d)f_\pi \int_{\Delta}^1 du  \frac{\varphi_\pi(u)}{u}e^{-\Xi}-
  \frac{(e_um_u^3-e_dm_d^3)f_\pi m_\pi^2}{4M^4} \int_{\Delta}^1 du \frac{A(u)}{u^3}e^{-\Xi}\nonumber\\
&&-e_uf_{3\pi}\int_0^1dv \int_0^1d\alpha_g
\int_0^{1-\alpha_g}d\alpha_u
\varphi_{3\pi}(\alpha_d,\alpha_g,\alpha_u)\Theta(u-\Delta)\nonumber\\
&&\left\{\frac{(1+2v) m_\pi^2
}{uM^2}-2(1-v)\frac{d}{du}\frac{1}{u}\right\}e^{-\Xi}\mid_{u=(1-v)\alpha_g+\alpha_u}
\nonumber\\
&&+e_df_{3\pi}\int_0^1dv \int_0^1d\alpha_g
\int_0^{1-\alpha_g}d\alpha_d
\varphi_{3\pi}(\alpha_d,\alpha_g,\alpha_u)\Theta(u-\Delta)\nonumber\\
&&\left\{\frac{(1+2v) m_\pi^2
}{uM^2}-2(1-v)\frac{d}{du}\frac{1}{u}\right\}e^{-\Xi}\mid_{u=(1-v)\alpha_g+\alpha_d}
\nonumber\\
&&+\frac{2f_\pi m_\pi^4}{M^4}\int_0^1dv v \int_0^1
d\alpha_g\int_0^{\alpha_g}
d\beta\int_0^{1-\beta}d\alpha \nonumber\\
&&\frac{e_um_u\Phi(1-\alpha-\beta,\beta,\alpha)+e_dm_d\widetilde{\Phi}(\alpha,\beta,1-\alpha-\beta)}
{u^2}\Theta(u-\Delta)e^{-\Xi}\mid_{u=1-v\alpha_g}
\nonumber \\
&& -\frac{2e_um_uf_\pi m_\pi^4}{M^4}\int_0^1 dv\int_0^1
d\alpha_g\int_0^{1-\alpha_g} d\alpha_u
 \int_0^{\alpha_u}d\alpha \nonumber\\
 &&\frac{\Phi(1-\alpha-\alpha_g,\alpha_g,\alpha)\Theta(u-\Delta)}
{u^2}e^{-\Xi}\mid_{u=(1-v)\alpha_g+\alpha_u}\nonumber\\
&& -\frac{2e_dm_df_\pi m_\pi^4}{M^4}\int_0^1 dv\int_0^1
d\alpha_g\int_0^{1-\alpha_g} d\alpha_d
 \int_0^{\alpha_d}d\alpha \nonumber\\
 &&\frac{\widetilde{\Phi}(\alpha,\alpha_g,1-\alpha-\alpha_g)\Theta(u-\Delta)}
{u^2}e^{-\Xi}\mid_{u=(1-v)\alpha_g+\alpha_d}\nonumber\\
&&+\frac{e_um_u f_\pi m_\pi^2}{M^2}  \int_0^1dv \int_0^1d\alpha_g
\int_0^{1-\alpha_g}d\alpha_u
\frac{\Psi(\alpha_d,\alpha_g,\alpha_u)\Theta(u-\Delta)}{u^2}e^{-\Xi}\mid_{u=(1-v)\alpha_g+\alpha_u}
 \nonumber\\
 &&+\frac{e_dm_d f_\pi m_\pi^2}{M^2}  \int_0^1dv \int_0^1d\alpha_g
\int_0^{1-\alpha_g}d\alpha_d
\frac{\widetilde{\Psi}(\alpha_d,\alpha_g,\alpha_u)\Theta(u-\Delta)}{u^2}e^{-\Xi}\mid_{u=(1-v)\alpha_g+\alpha_d}
\, ,\nonumber\\
\end{eqnarray}

\begin{eqnarray}
&&\frac{f_\pi m_\pi^2}{m_u+m_d}F_\pi^q(q^2)e^{-\frac{m_\pi^2}{M^2}}
\nonumber\\
  &=&  \frac{f_\pi
m_\pi^2}{m_u+m_d}\int_{\Delta}^1du
\frac{\varphi_p(u)}{u}e^{-\Xi}-\frac{(e_um_u-e_dm_d)f_\pi
m_\pi^2}{M^2}\int_{\Delta}^1du \int_0^u
dt\frac{B(t)}{u^2}e^{-\Xi} \nonumber\\
&&-\frac{1}{6}\frac{f_\pi m_\pi^2}{m_u+m_d}\int_{\Delta}^1du
\varphi_\sigma(u)\frac{d}{du} \frac{1}{u}e^{-\Xi}
  \nonumber\\
  &&-\frac{e_uf_{3\pi}m_\pi^2}{M^2}\int_0^1dv \int_0^1d\alpha_g \int_0^{1-\alpha_g}d\alpha_u
\varphi_{3\pi}(\alpha_d,\alpha_g,\alpha_u)\nonumber\\
&&\Theta(u-\Delta)\frac{1+2v
}{u^2}e^{-\Xi}\mid_{u=(1-v)\alpha_g+\alpha_u}
\nonumber\\
&&+\frac{e_df_{3\pi}m_\pi^2}{M^2}\int_0^1dv \int_0^1d\alpha_g
\int_0^{1-\alpha_g}d\alpha_d
\varphi_{3\pi}(\alpha_d,\alpha_g,\alpha_u)\nonumber\\
&&\Theta(u-\Delta)\frac{1+2v
}{u^2}e^{-\Xi}\mid_{u=(1-v)\alpha_g+\alpha_d}
\nonumber\\
&&+\frac{2f_\pi m_\pi^4}{M^4}\int_0^1dv v \int_0^1
d\alpha_g\int_0^{\alpha_g}
d\beta\int_0^{1-\beta}d\alpha \nonumber \\
&&
\frac{e_um_u\Phi(1-\alpha-\beta,\beta,\alpha)+e_dm_d\widetilde{\Phi}(\alpha,\beta,1-\alpha-\beta)
}{u^3}\Theta(u-\Delta)e^{-\Xi}\mid_{1-v\alpha_g}
\nonumber \\
&& -\frac{2e_um_uf_\pi m_\pi^4}{M^4}\int_0^1 dv\int_0^1
d\alpha_g\int_0^{1-\alpha_g} d\alpha_u
 \int_0^{\alpha_u}d\alpha \nonumber\\
 &&\frac{\Phi(1-\alpha-\alpha_g,\alpha_g,\alpha)}{u^3}\Theta(u-\Delta)e^{-\Xi}\mid_{u=(1-v)\alpha_g+\alpha_u}
\nonumber\\
&& -\frac{2e_dm_df_\pi m_\pi^4}{M^4}\int_0^1 dv\int_0^1
d\alpha_g\int_0^{1-\alpha_g} d\alpha_d
 \int_0^{\alpha_d}d\alpha \nonumber\\
 &&\frac{\Phi(\alpha,\alpha_g,1-\alpha-\alpha_g)}{u^3}\Theta(u-\Delta)e^{-\Xi}\mid_{u=(1-v)\alpha_g+\alpha_d}
\, ,
\end{eqnarray}

where
\begin{eqnarray}
\Xi &=& \frac{m_q^2+u(1-u)m_\pi^2-(1-u)q^2}{uM^2} \, , \nonumber\\
\Delta&=&\frac{m_q^2-q^2}{s_0-q^2} \, ,\nonumber\\
\Theta(x)&=&1  \,\, \rm{for\,\, x\geq 0} \, ,
\end{eqnarray}
 and    the  $s_0$ is threshold parameter.

\section{Numerical result and discussion}
The input parameters of the light-cone distribution amplitudes are
taken as  $\lambda_3=0.0$, $f_{3\pi}=(0.45\pm0.15)\times
10^{-2}\,\,\rm{GeV}^2$, $\omega_3=-1.5\pm0.7$, $\omega_4=0.2\pm0.1$,
 $a_1=0.0 $, $a_2=0.25\pm 0.15$, $a_4=0.0$, $\eta_4=10.0\pm3.0$
\cite{LCSR1,LCSR2,LCSR3,LCSR4,LCSR5,Belyaev94,PSLC1,PSLC2,PSLC3,PSLC4};
and $m_u=m_d=m_q=(5.6\pm 1.6) \,\,\rm{MeV}$,
$f_\pi=0.130\,\,\rm{GeV}$, $m_{\pi} =135\,\,\rm{MeV}$. The threshold
parameter is chosen to be $s_0=0.8\,\,\rm{GeV}^2$, which can
reproduce the value of the decay constant
$f_{\pi}=0.130\,\,\rm{GeV}$ in the QCD sum rules.

In this article, we take the values of the coefficients $a_i$ of the
twist-2 light-cone distribution amplitude $\varphi_\pi(u)$ from the
conventional QCD sum rules \cite{PSLC1,PSLC4}. The $\varphi_\pi(u)$
has been  analyzed with the light-cone QCD sum rules and (non-local
condensates) QCD sum rules confronting with the high precision CLEO
data on the $  \gamma \gamma^* \to \pi^0$ transition form-factor
\cite{Schmedding1999-24,Bakulev2001-24,Bakulev2002-24,Bakulev2003-24,Bakulev2006-24,Bakulev2006-24-2}.
We also study the electromagnetic form-factors $F_\pi^p(Q^2)$ and
$F_\pi^q(Q^2)$ with the values $a_2=0.29$ and $a_4=-0.21$ at $\mu=1$
GeV, which are obtained via  one-loop renormalization group equation
for the central  values $a_2=0.268$ and $a_4=-0.186$ at
$\mu^2=1.35\,\,\rm{GeV}^2$  from  the (non-local condensates) QCD
sum rules with improved model \cite{Bakulev2006-24-2}.

The Borel parameters in the two sum rules  are taken as
$M^2=(0.8-1.5)\,\, \rm{GeV}^2$, in this region, the values of the
electromagnetic  form-factors $F_\pi^p(Q^2)$ and $F_\pi^q(Q^2)$ are
rather stable. In this article, we take the special value
$M^2=1.0\,\,\rm{GeV}^2$ in numerical calculations, although such a
definite Borel parameter cannot take into account some
uncertainties, the predictive power cannot be impaired
qualitatively.
\begin{figure}
\centering
  \includegraphics[totalheight=7cm,width=7cm]{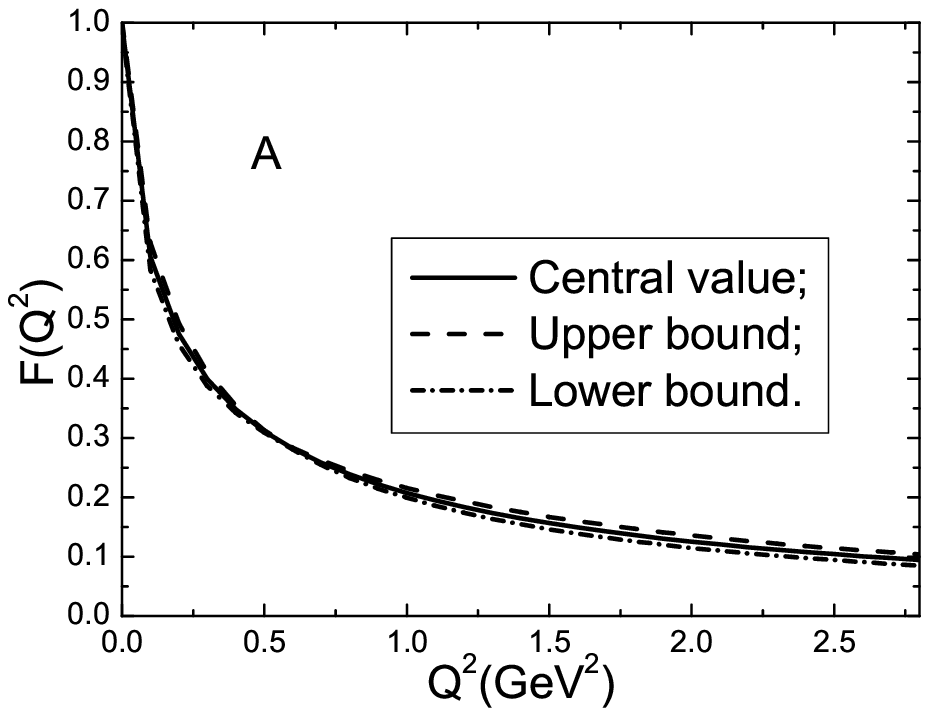}
  \includegraphics[totalheight=7cm,width=7cm]{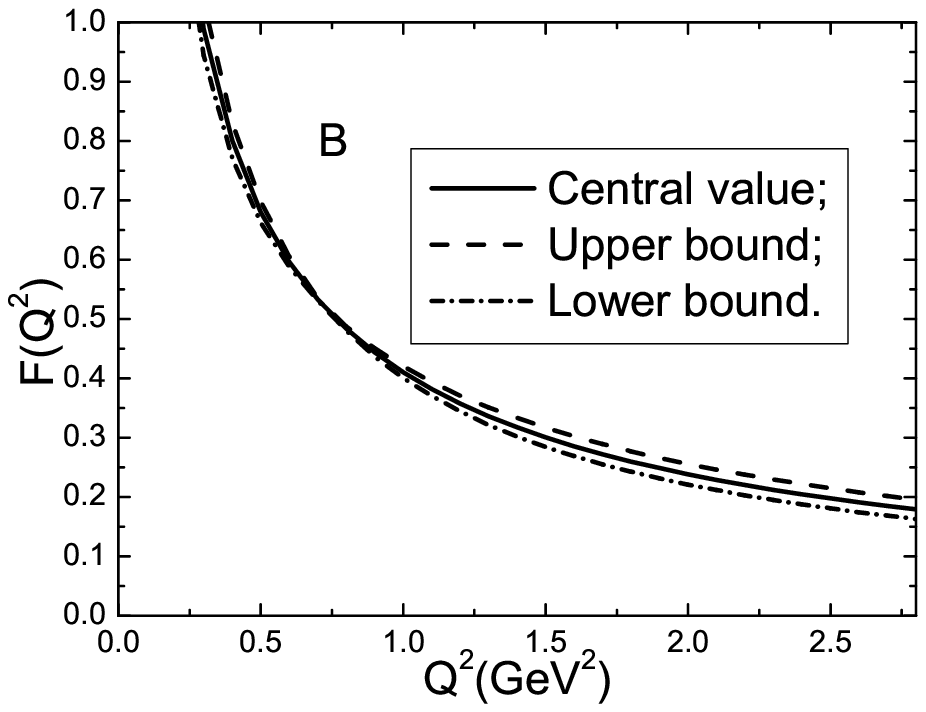}
   \caption{The   $F_\pi^{p}(Q^2)$(A) and $F_\pi^{q}(Q^2)$(B) with the parameter $M^2=1\,\,\rm{GeV^2}$. }
\end{figure}

\begin{figure}
\centering
  \includegraphics[totalheight=8cm,width=9cm]{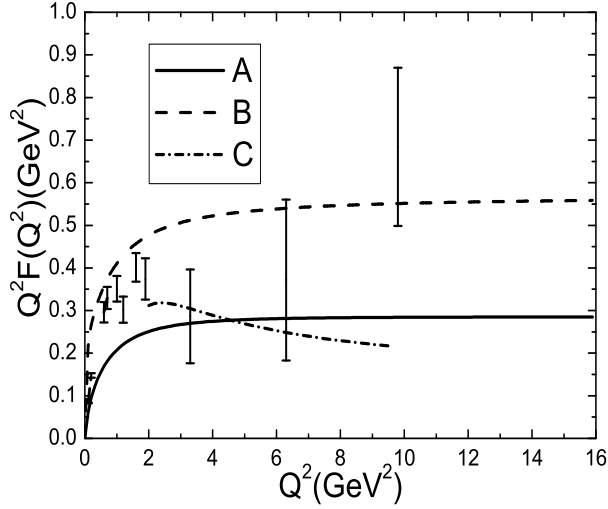}
     \caption{The numerical values of the form-factors   $Q^2F_\pi^{p}(Q^2)$(A) and $Q^2F_\pi^{q}(Q^2)$(B) in
     comparison with the experimental data \cite{exp3,exp6,exp8},
     line $C$ corresponds to the central values of $Q^2F_\pi(Q^2)$ from the light-cone sum rules with the axial-vector current
     \cite{Bijnens02}.         }
\end{figure}

In the two sum rules in Eqs.(7-8),  the dominant  contributions come
from the two-particle twist-3 light-cone distribution amplitudes
$\varphi_p(u)$ and $\varphi_\sigma(u)$ due to the pseudoscalar
current $J_\pi(x)$. The different values of the coefficients of the
$\varphi_\pi(u)$  obtained  in Ref.\cite{PSLC4} and
Ref.\cite{Bakulev2006-24-2} respectively can lead to results of
minor difference.  If we choose the axial-vector current to
interpolate the $\pi$ meson, the main contributions come from the
twist-2 light-cone distribution amplitude $\varphi_\pi(u)$
\cite{Brodsky80,HnLi921,HnLi922,Huang04,Bakulev04}. The
uncertainties concerning  the denominator $\frac{1}{m_u+m_d}$ are
canceled out with each other, see Eqs.(7-8), which result in small
net uncertainties.

Taking into account all the uncertainties, finally we obtain the
numerical values of the electromagnetic form-factors
$F_\pi^{p}(Q^2)$ and $F_\pi^{q}(Q^2)$, which are shown in the
Figs.1-2, at zero momentum transfer,
\begin{eqnarray}
 F_\pi^{p}(0) &=&0.999\pm 0.001    \, , \nonumber \\
 F_\pi^{q}(0) &=&16.05\pm 1.82  \, ,
 \end{eqnarray}
the  parameters of the twist-2 light-cone distribution amplitude
 $\varphi_\pi(u)$ obtained in Ref.\cite{Bakulev2006-24-2} can reduce
 the values of the form-factors $F_\pi^p(Q^2)$ and $F_\pi^q(Q^2)$
slightly, about $(1-2)\%$.

 Comparing the
experimental data (extrapolated to the limit $Q^2\rightarrow 0$, or
the normalization condition $F_\pi(0)=1$)
\cite{exp1,exp2,exp3,exp4,exp5,exp6,exp8} and theoretical estimation
with the vector meson dominance theory \cite{VMD}, our numerical
value $F_\pi^{p}(0) =0.999\pm 0.001$ is excellent.  The value
$F_\pi^{q}(0) =16.05\pm 1.82$ is too large to make any reliable
prediction, however, it is not un-expected. From the two sum rules,
we can see that the terms of the $F_\pi^q(Q^2)$ are companied with
 an extra factor $\frac{1}{u}$, for example,
\begin{eqnarray}
F_\pi^{q}(Q^2) &\propto& \int_{\Delta}^1du
\frac{\varphi_p(u)}{u}e^{-\frac{m_q^2+u(1-u)m_\pi^2-(1-u)q^2}{uM^2}}=
\int_{\Delta}^1
\frac{du}{u}e^{-\frac{m_q^2+u(1-u)m_\pi^2-(1-u)q^2}{uM^2}} \,
,\nonumber \\
F_\pi^{p}(Q^2) &\propto& \int_{\Delta}^1du
\varphi_p(u)e^{-\frac{m_q^2+u(1-u)m_\pi^2-(1-u)q^2}{uM^2}}=
\int_{\Delta}^1 du e^{-\frac{m_q^2+u(1-u)m_\pi^2-(1-u)q^2}{uM^2}} \,
,\nonumber
\end{eqnarray}
where we have taken  the asymptotic distribution amplitude
$\varphi_p(u)=1$. The value of the $F_\pi^{q}(Q^2)$ is greatly
enhanced in the region of small-$Q^2$ due to the extra
$\frac{1}{u}$, in the limit $Q^2=0$, $\Delta\approx 0.00004$, the
dominant contributions come from the end-point of the light-cone
distribution amplitudes. We should introduce extra phenomenological
form-factors (for example, the Sudakov factor
\cite{HnLi921,HnLi922}) to suppress the contribution from the
end-point. The value of the $F_\pi^{p}(Q^2)$ is more reliable at
small momentum transfers.

In the light-cone QCD sum rules, we carry out the operator product
expansion near the light-cone $x^2\approx 0$, which corresponds to
 $Q^2\gg 0$ and $P^2 \gg 0 $, the two sum rules for the form-factors
  $F_\pi^p(Q^2)$ and $F_\pi^q(Q^2)$ are valid at large momentum transfers.
  We take  the analytical expressions of the $F_\pi^p(Q^2)$ and $F_\pi^q(Q^2)$ in Eqs.(7-8)
  as some functions which model the electromagnetic
form-factor $ F_\pi(Q^2) $   at large momentum transfers,   then
extrapolate the $ F_\pi^{p}(Q^2) $ and $F_\pi^{q} (Q^2)$ to  zero
momentum transfer  (or beyond zero momentum transfer) with
analytical continuation in hope of obtaining some interesting
results \footnote{We can borrow some ideas from the electromagnetic
$\pi$-photon form-factor $f_{\gamma^* \pi^0 }(Q^2)$. The value of
 $f_{\gamma^* \pi^0 }(0)$ is fixed by partial conservation of
 the axial current and the effective  anomaly lagrangian,
$f_{\gamma^* \pi^0 }(0) = \frac1{\pi f_{\pi}}$. In the limit of
large-$Q^2$,   perturbative QCD predicts that $f_{\gamma^* \pi^0
}(Q^2) =\frac{4\pi f_{\pi}}{Q^2} $.
 The Brodsky-Lepage interpolation formula \cite{BJ81}
 \begin{eqnarray}
f_{\gamma^* \pi^0 }(Q^2) = \frac{1}{ \pi f_{\pi} \left [1+Q^2/(4
\pi^2 f_{\pi}^2) \right ]} =\frac{1}{ \pi f_\pi (1+Q^2/s_0) }
\nonumber
\end{eqnarray}
 can reproduce both the value at $Q^2 =0$
  and the  behavior at large-$Q^2$. The energy scale $s_0$ ($s_0 = 4 \pi^2
f_{\pi}^2 \approx 0.67 \,\, \rm{GeV}^2$) is numerically   close to
the squared mass of the $\rho$ meson, $m_{\rho}^2 \approx 0.6 \,\,
\rm{GeV}^2$. The Brodsky-Lepage interpolation formula is similar to
the result of the vector meson dominance approach, $f_{\gamma^*
\pi^0 }(Q^2) = 1/\left\{\pi f_\pi (1+Q^2/m_{\rho}^2)\right\}$. In
the latter case, the calculation is performed at the timelike energy
scale $q^2<1\,\,\rm{GeV}^2$ and the electromagnetic  current is
saturated by the vector meson $\rho$, where the mass $m_{\rho}$
serves as a parameter determining the pion charge radius. With a
slight modification of the mass parameter,
$m_\rho=\Lambda_\pi=776\,\,\rm{MeV}$, the experimental data can be
well described by the single-pole formula at the interval
$Q^2=(0-10)\,\,\rm{GeV}^2$ \cite{CLEO97}. In Ref.\cite{Wang063}, the
four form-factors of  $\Sigma \to n$ have satisfactory behaviors at
large $Q^2$, which are expected by naive power counting rules, and
they have finite values at $Q^2=0$. The analytical expressions of
the four form-factors $f_1(Q^2)$, $f_2(Q^2)$, $g_1(Q^2)$ and
$g_2(Q^2)$ are taken as  Brodsky-Lepage type of interpolation
formulae, although they are calculated at rather large $Q^2$, the
extrapolation to lower energy transfer has no solid theoretical
foundation. The numerical values of $f_1(0)$, $f_2(0)$, $g_1(0)$ and
$g_2(0)$ are
 compatible with the  experimental data and  theoretical
 calculations (in magnitude).
 In Ref.\cite{Wang071}, the vector form-factors $f_{K\pi}^+(Q^2)$
 and  $f_{K\pi}^-(Q^2)$ are
 also taken as Brodsky-Lepage  type of interpolation formulae, the behaviors of low momentum
 transfer  are rather good in some channels.}. It is obvious that the model
functions $F_\pi^p(Q^2)$ and $F_\pi^q(Q^2)$ may
  have good or bad low-$Q^2$ behaviors, although
  they have solid theoretical foundation at large momentum
transfers. We extrapolate the model functions tentatively to zero
momentum transfer, systematic errors maybe very large and the
results maybe unreliable. The predictions merely  indicate the
possible values of the light-cone QCD sum rules approach, they
should be confronted with the experimental data or other theoretical
approaches. The numerical results indicate that the small-$Q^2$
behavior of the $F_\pi^p(Q^2)$ is better than that  of the
$F_\pi^q(Q^2)$, so we take the value of the $F_\pi^p(Q^2)$  at $Q^2<
1\,\,\rm{GeV}^2$.

The electromagnetic form-factors $F_\pi^p(Q^2)$ and $F_\pi^q(Q^2)$
are complex functions of the input parameters, in principle, they
can be expanded in terms of Taylor series of $\frac{1}{Q^2}$ for
large-$Q^2$. At large momentum transfer, for example,
$Q^2=(6-16)\,\,\rm{GeV}^2$, the central values of the two
form-factors $F_\pi^p(Q^2)$ and $F_\pi^q(Q^2)$ can be fitted
numerically as
\begin{eqnarray}
F_\pi^{p}(Q^2)&=& \frac{0.285}{Q^2} \, ,\nonumber \\
F_\pi^{q}(Q^2)&=&\frac{0.554}{Q^2} \,  ,
\end{eqnarray}
which are comparable with the experimental data
\cite{exp1,exp2,exp3,exp4,exp5,exp6,exp8} and theoretical
estimations, for examples, the light-cone QCD sum rules
\cite{Braun94,Braun00,Bijnens02,Agaev05}, perturbative QCD
\cite{Brodsky80,HnLi921,HnLi922,Huang04,Bakulev04}. In Fig.2, we
plot the electromagnetic form-factor $Q^2F_\pi(Q^2)$ comparing with
the experimental data in Refs. \cite{exp3,exp6,exp8} and prediction
of the light-cone QCD sum rules with the axial-vector current in
Ref.\cite{Bijnens02}. For more literatures, one can consult
Ref.\cite{Exp06PhD}.

The large-$Q^2$ behavior $F_\pi(Q^2)\sim\frac{1}{Q^2}$ is expected
from the naive power counting rules \cite{BJ811,BJ812,BJ813}. At
large-$Q^2$, the  $i$-th term in the form-factors $F_\pi^p(Q^2)$ and
$F_\pi^q(Q^2)$ respectively can be expanded as
$\frac{A_i}{Q^2}+\frac{B_i}{Q^4}+\frac{C_i}{Q^6}+\cdots$,  the terms
proportional to $\frac{1}{Q^{2n}}$ with $n\geq 2$ are canceled out
approximately with each other, i.e. $\sum B_i\approx0$, $\sum
C_i\approx0$, $\cdots$.  Finally we obtain $\sum A_i=0.285$ and
$\sum A_i=0.554$ for the $F_\pi^p(Q^2)$ and $F_\pi^q(Q^2)$
respectively. Due to partial conservation of the axial-vector
current, the axial-vector current has no-vanishing coupling with the
$\pi$ meson, we can choose either the axial-vector current or the
pseudoscalar current to interpolate the $\pi$ meson. They  can lead
to different sum rules, in the case of the axial-vector current, the
soft contributions proportional to $\frac{1}{Q^{4}}$ manifest
themselves at large-$Q^2$ \cite{Braun00,Bijnens02}, see Fig.2,  more
experimental data are needed  to select the pertinent sum rules.

In the limit of large-$Q^2$, $F_\pi(Q^2)\sim\frac{1}{Q^2}$, which is
consistent with the prediction of perturbative QCD theory, i.e.
hard-gluon exchange between the $u$ and $d$ quarks dominates over
Feynman mechanism.

\section{Conclusion}

In this article, we calculate the electromagnetic form-factor of the
$\pi$ meson with the light-cone QCD sum rules.  Our numerical value
$F_\pi^{p}(0) =0.999\pm 0.001$ is in excellent agreement with the
experimental data (extrapolated to the limit $Q^2\rightarrow 0$ or
the normalization condition $F_\pi(0)=1$). For large momentum
transfers, the values from the two sum rules are all comparable with
the experimental data and theoretical estimations.

\section*{Appendix}
The explicit expressions of the correlation functions,
\begin{eqnarray}
\Pi_p^u&=&  \frac{f_\pi m_\pi^2}{m_u+m_d}\int_0^1du
\frac{u\varphi_p(u)}{m_u^2-(q+up)^2}-m_uf_\pi m_\pi^2\int_0^1du
\int_0^u
dt\frac{u B(t)}{\left[m_u^2-(q+up)^2\right]^2} \nonumber\\
&&+\frac{1}{6}\frac{f_\pi m_\pi^2}{m_u+m_d}\int_0^1du
\varphi_\sigma(u)\left\{\left[1-u\frac{d}{du}
\right]\frac{1}{m_u^2-(q+up)^2}+\frac{2m_u^2}{\left[m_u^2-(q+up)^2\right]^2}\right\}
  \nonumber\\
  &&+m_uf_\pi \int_0^1 du \left\{ \frac{\varphi_\pi(u)}{m_u^2-(q+up)^2} -\frac{m_\pi^2m_u^2}{2} \frac{A(u)}{\left[m_u^2-(q+up)^2\right]^3}\right\}\nonumber\\
&&-f_{3\pi}\int_0^1dv \int_0^1d\alpha_g \int_0^{1-\alpha_g}d\alpha_u
\varphi_{3\pi}(\alpha_d,\alpha_g,\alpha_u)\nonumber\\
&&\left\{\frac{(1+2v)u m_\pi^2
}{\left[m_u^2-(q+up)^2\right]^2}-2(1-v)\frac{d}{du}\frac{1}{m_u^2-(q+up)^2}\right\}\mid_{u=(1-v)\alpha_g+\alpha_u}
\nonumber\\
&&+4m_uf_\pi m_\pi^4\int_0^1dv v \int_0^1 d\alpha_g\int_0^{\alpha_g}
d\beta\int_0^{1-\beta}d\alpha
\frac{u\Phi(1-\alpha-\beta,\beta,\alpha)}{\left[m_u^2-(q+up)^2\right]^3}\mid_{1-v\alpha_g}
\nonumber \\
&& -4m_uf_\pi m_\pi^4\int_0^1 dv\int_0^1
d\alpha_g\int_0^{1-\alpha_g} d\alpha_u
 \int_0^{\alpha_u}d\alpha
\frac{u\Phi(1-\alpha-\alpha_g,\alpha_g,\alpha)}
{\left[m_u^2-(q+up)^2\right]^3}\mid_{u=(1-v)\alpha_g+\alpha_u}\nonumber\\
&&+m_u f_\pi m_\pi^2  \int_0^1dv \int_0^1d\alpha_g
\int_0^{1-\alpha_g}d\alpha_u
\frac{\Psi(\alpha_d,\alpha_g,\alpha_u)}{\left[m_u^2-(q+up)^2\right]^2}\mid_{u=(1-v)\alpha_g+\alpha_u}
\, ,
\end{eqnarray}

\begin{eqnarray}
\Pi_p^d&=& - \frac{f_\pi m_\pi^2}{m_u+m_d}\int_0^1du
\frac{u\varphi_p(u)}{m_d^2-(q+up)^2}+m_df_\pi m_\pi^2\int_0^1du
\int_0^u
dt\frac{u B(t)}{\left[m_d^2-(q+up)^2\right]^2} \nonumber\\
&&-\frac{1}{6}\frac{f_\pi m_\pi^2}{m_u+m_d}\int_0^1du
\varphi_\sigma(u)\left\{\left[1-u\frac{d}{du}
\right]\frac{1}{m_d^2-(q+up)^2}+\frac{2m_d^2}{\left[m_d^2-(q+up)^2\right]^2}\right\}
  \nonumber\\
  &&-m_df_\pi \int_0^1 du \left\{ \frac{\varphi_\pi(u)}{m_d^2-(q+up)^2} -\frac{m_\pi^2m_d^2}{2} \frac{A(u)}{\left[m_d^2-(q+up)^2\right]^3}\right\}\nonumber\\
&&+f_{3\pi}\int_0^1dv \int_0^1d\alpha_g \int_0^{1-\alpha_g}d\alpha_d
\varphi_{3\pi}(\alpha_d,\alpha_g,\alpha_u)\nonumber\\
&&\left\{\frac{(1+2v)u m_\pi^2
}{\left[m_d^2-(q+up)^2\right]^2}-2(1-v)\frac{d}{du}\frac{1}{m_d^2-(q+up)^2}\right\}\mid_{u=(1-v)\alpha_g+\alpha_d}
\nonumber\\
&&+4m_df_\pi m_\pi^4\int_0^1dv v \int_0^1 d\alpha_g\int_0^{\alpha_g}
d\beta\int_0^{1-\beta}d\alpha
\frac{u\widetilde{\Phi}(\alpha,\beta,1-\alpha-\beta)}{\left[m_d^2-(q+up)^2\right]^3}\mid_{1-v\alpha_g}
\nonumber \\
&& -4m_df_\pi m_\pi^4\int_0^1 dv\int_0^1
d\alpha_g\int_0^{1-\alpha_g} d\alpha_d
 \int_0^{\alpha_d}d\alpha
\frac{u\widetilde{\Phi}(\alpha,\alpha_g,1-\alpha-\alpha_g)}
{\left[m_d^2-(q+up)^2\right]^3}\mid_{u=(1-v)\alpha_g+\alpha_d}\nonumber\\
&&+m_d f_\pi m_\pi^2  \int_0^1dv \int_0^1d\alpha_g
\int_0^{1-\alpha_g}d\alpha_d
\frac{\widetilde{\Psi}(\alpha_d,\alpha_g,\alpha_u)}{\left[m_d^2-(q+up)^2\right]^2}\mid_{u=(1-v)\alpha_g+\alpha_d}
\, ,
\end{eqnarray}

\begin{eqnarray}
\Pi_q^u&=&  \frac{f_\pi m_\pi^2}{m_u+m_d}\int_0^1du
\frac{\varphi_p(u)}{m_u^2-(q+up)^2}-m_uf_\pi m_\pi^2\int_0^1du
\int_0^u
dt\frac{B(t)}{\left[m_u^2-(q+up)^2\right]^2} \nonumber\\
&&-\frac{1}{6}\frac{f_\pi m_\pi^2}{m_u+m_d}\int_0^1du
\varphi_\sigma(u)\frac{d}{du} \frac{1}{m_u^2-(q+up)^2}
  \nonumber\\
  &&-f_{3\pi}m_\pi^2\int_0^1dv \int_0^1d\alpha_g \int_0^{1-\alpha_g}d\alpha_u
\varphi_{3\pi}(\alpha_d,\alpha_g,\alpha_u)\frac{1+2v
}{\left[m_u^2-(q+up)^2\right]^2}\mid_{u=(1-v)\alpha_g+\alpha_u}
\nonumber\\
&&+4m_uf_\pi m_\pi^4\int_0^1dv v \int_0^1 d\alpha_g\int_0^{\alpha_g}
d\beta\int_0^{1-\beta}d\alpha
\frac{\Phi(1-\alpha-\beta,\beta,\alpha)}{\left[m_u^2-(q+up)^2\right]^3}\mid_{1-v\alpha_g}
\nonumber \\
&& -4m_uf_\pi m_\pi^4\int_0^1 dv\int_0^1
d\alpha_g\int_0^{1-\alpha_g} d\alpha_u
 \int_0^{\alpha_u}d\alpha
\frac{\Phi(1-\alpha-\alpha_g,\alpha_g,\alpha)}{\left[m_u^2-(q+up)^2\right]^3}\mid_{u=(1-v)\alpha_g+\alpha_u}
 \, ,\nonumber \\
\end{eqnarray}

\begin{eqnarray}
\Pi_q^d&=& - \frac{f_\pi m_\pi^2}{m_u+m_d}\int_0^1du
\frac{\varphi_p(u)}{m_d^2-(q+up)^2}+m_df_\pi m_\pi^2\int_0^1du
\int_0^u
dt\frac{B(t)}{\left[m_d^2-(q+up)^2\right]^2} \nonumber\\
&&+\frac{1}{6}\frac{f_\pi m_\pi^2}{m_u+m_d}\int_0^1du
\varphi_\sigma(u)\frac{d}{du} \frac{1}{m_d^2-(q+up)^2}
  \nonumber\\
  &&+f_{3\pi}m_\pi^2\int_0^1dv \int_0^1d\alpha_g
  \int_0^{1-\alpha_g}d\alpha_d
\varphi_{3\pi}(\alpha_d,\alpha_g,\alpha_u)\frac{1+2v
}{\left[m_d^2-(q+up)^2\right]^2}\mid_{u=(1-v)\alpha_g+\alpha_d}
\nonumber\\
&&+4m_df_\pi m_\pi^4\int_0^1dv v \int_0^1 d\alpha_g\int_0^{\alpha_g}
d\beta\int_0^{1-\beta}d\alpha
\frac{\widetilde{\Phi}(\alpha,\beta,1-\alpha-\beta)}{\left[m_d^2-(q+up)^2\right]^3}\mid_{1-v\alpha_g}
\nonumber \\
&& -4m_df_\pi m_\pi^4\int_0^1 dv\int_0^1
d\alpha_g\int_0^{1-\alpha_g} d\alpha_d
 \int_0^{\alpha_d}d\alpha
\frac{\widetilde{\Phi}(\alpha,\alpha_g,1-\alpha-\alpha_g)}{\left[m_d^2-(q+up)^2\right]^3}\mid_{u=(1-v)\alpha_g+\alpha_d}
 \, ,\nonumber \\
\end{eqnarray}

where
\begin{eqnarray}
\Phi&=&A_\parallel+A_\perp-V_\perp-V_\parallel \,, \nonumber \\
 \widetilde{\Phi}&=&A_\parallel+A_\perp+V_\perp+V_\parallel  \,, \nonumber \\
\Psi&=&2A_\perp-2V_\perp -A_\parallel+V_\parallel \,, \nonumber \\
\widetilde{\Psi}&=&2A_\perp+2V_\perp -A_\parallel-V_\parallel\,.
\nonumber
\end{eqnarray}

The light-cone distribution amplitudes of the $\pi$ meson are
defined as
\begin{eqnarray}
\langle0| {\bar u} (0) \gamma_\mu \gamma_5 d(x) |\pi(p)\rangle& =& i
f_\pi p_\mu \int_0^1 du  e^{-i u p\cdot x}
\left\{\varphi_\pi(u)+\frac{m_\pi^2x^2}{16}
A(u)\right\}\nonumber\\
&&+if_\pi m_\pi^2\frac{x_\mu}{2p\cdot x}
\int_0^1 du  e^{-i u p \cdot x} B(u) \, , \nonumber\\
\langle0| {\bar u} (0) i \gamma_5 d(x) |\pi(p)\rangle &=&
\frac{f_\pi m_\pi^2}{ m_u+m_d}
\int_0^1 du  e^{-i u p \cdot x} \varphi_p(u)  \, ,  \nonumber\\
\langle0| {\bar u} (0) \sigma_{\mu \nu} \gamma_5 d(x) |\pi(p)\rangle
&=&i(p_\mu x_\nu-p_\nu x_\mu)  \frac{f_\pi m_\pi^2}{6 (m_u+m_d)}
\int_0^1 du
e^{-i u p \cdot x} \varphi_\sigma(u) \, ,  \nonumber\\
\langle0| {\bar u} (0) \sigma_{\alpha \beta} \gamma_5 g_s G_{\mu
\nu}(v x)d(x) |\pi(p)\rangle&=& f_{3 \pi}\left\{(p_\mu p_\alpha
g^\bot_{\nu
\beta}-p_\nu p_\alpha g^\bot_{\mu \beta}) -(p_\mu p_\beta g^\bot_{\nu \alpha}\right.\nonumber\\
&&\left.-p_\nu p_\beta g^\bot_{\mu \alpha})\right\} \int {\cal
D}\alpha_i \varphi_{3 \pi} (\alpha_i)
e^{-ip \cdot x(\alpha_d+v \alpha_g)} \, ,\nonumber\\
\langle0| {\bar u} (0) \gamma_{\mu} \gamma_5 g_s G_{\alpha
\beta}(vx)d(x) |\pi(p)\rangle&=&  p_\mu  \frac{p_\alpha
x_\beta-p_\beta x_\alpha}{p
\cdot x}f_\pi m_\pi^2\nonumber\\
&&\int{\cal D}\alpha_i A_{\parallel}(\alpha_i) e^{-ip\cdot
x(\alpha_d +v \alpha_g)}\nonumber \\
&&+ f_\pi m_\pi^2 (p_\beta g_{\alpha\mu}-p_\alpha
g_{\beta\mu})\nonumber\\
&&\int{\cal D}\alpha_i A_{\perp}(\alpha_i)
e^{-ip\cdot x(\alpha_d +v \alpha_g)} \, ,  \nonumber\\
\langle0| {\bar u} (0) \gamma_{\mu}  g_s \tilde G_{\alpha
\beta}(vx)d(x) |\pi(p)\rangle&=& p_\mu  \frac{p_\alpha
x_\beta-p_\beta x_\alpha}{p \cdot
x}f_\pi m_\pi^2\nonumber\\
&&\int{\cal D}\alpha_i V_{\parallel}(\alpha_i) e^{-ip\cdot
x(\alpha_d +v \alpha_g)}\nonumber \\
&&+ f_\pi m_\pi^2 (p_\beta g_{\alpha\mu}-p_\alpha
g_{\beta\mu})\nonumber\\
&&\int{\cal D}\alpha_i V_{\perp}(\alpha_i) e^{-ip\cdot x(\alpha_d +v
\alpha_g)} \, ,
\end{eqnarray}
where  $\tilde G_{\alpha \beta}= {1\over 2} \epsilon_{\alpha \beta
\mu\nu} G^{\mu\nu} $ and ${\cal{D}} \alpha_i =d \alpha_1 d \alpha_2
d \alpha_3 \delta(1-\alpha_1 -\alpha_2 -\alpha_3)$.

The  light-cone distribution amplitudes are parameterized as
\begin{eqnarray}
\varphi_\pi(u)&=&6u(1-u)
\left\{1+a_1C^{\frac{3}{2}}_1(2u-1)+a_2C^{\frac{3}{2}}_2(2u-1)+a_4C^{\frac{3}{2}}_4(2u-1)
\right\}\, , \nonumber\\
\varphi_p(u)&=&1+\left\{30\eta_3-\frac{5}{2}\rho^2\right\}C_2^{\frac{1}{2}}(2u-1)\nonumber \\
&&+\left\{-3\eta_3\omega_3-\frac{27}{20}\rho^2-\frac{81}{10}\rho^2 a_2\right\}C_4^{\frac{1}{2}}(2u-1)\, ,  \nonumber \\
\varphi_\sigma(u)&=&6u(1-u)\left\{1
+\left[5\eta_3-\frac{1}{2}\eta_3\omega_3-\frac{7}{20}\rho^2-\frac{3}{5}\rho^2 a_2\right]C_2^{\frac{3}{2}}(2u-1)\right\}\, , \nonumber \\
\varphi_{3\pi}(\alpha_i) &=& 360 \alpha_u \alpha_d \alpha_g^2 \left
\{1 +\lambda_3(\alpha_u-\alpha_d)+ \omega_3 \frac{1}{2} ( 7 \alpha_g
- 3) \right\} \, , \nonumber\\
V_{\parallel}(\alpha_i) &=& 120\alpha_u \alpha_d \alpha_g \left(
v_{00}+v_{10}(3\alpha_g-1)\right)\, ,
\nonumber \\
A_{\parallel}(\alpha_i) &=& 120 \alpha_u \alpha_d \alpha_g a_{10}
(\alpha_d-\alpha_u)\, ,
\nonumber\\
V_{\perp}(\alpha_i) &=& -30\alpha_g^2
\left\{h_{00}(1-\alpha_g)+h_{01}\left[\alpha_g(1-\alpha_g)-6\alpha_u
\alpha_d\right] \right.  \nonumber\\
&&\left. +h_{10}\left[
\alpha_g(1-\alpha_g)-\frac{3}{2}\left(\alpha_u^2+\alpha_d^2\right)\right]\right\}\,
, \nonumber\\
A_{\perp}(\alpha_i) &=&  30 \alpha_g^2 (\alpha_u-\alpha_d) \left\{h_{00}+h_{01}\alpha_g+\frac{1}{2}h_{10}(5\alpha_g-3)  \right\}, \nonumber\\
A(u)&=&6u(1-u)\left\{
\frac{16}{15}+\frac{24}{35}a_2+20\eta_3+\frac{20}{9}\eta_4 \right.
\nonumber \\
&&+\left[
-\frac{1}{15}+\frac{1}{16}-\frac{7}{27}\eta_3\omega_3-\frac{10}{27}\eta_4\right]C^{\frac{3}{2}}_2(2u-1)
\nonumber\\
&&\left.+\left[
-\frac{11}{210}a_2-\frac{4}{135}\eta_3\omega_3\right]C^{\frac{3}{2}}_4(2u-1)\right\}+\left\{
 -\frac{18}{5}a_2+21\eta_4\omega_4\right\} \nonumber\\
 && \left\{2u^3(10-15u+6u^2) \log u+2\bar{u}^3(10-15\bar{u}+6\bar{u}^2) \log \bar{u}
 \right. \nonumber\\
 &&\left. +u\bar{u}(2+13u\bar{u})\right\} \, ,\nonumber\\
 g_\pi(u)&=&1+g_2C^{\frac{1}{2}}_2(2u-1)+g_4C^{\frac{1}{2}}_4(2u-1)\, ,\nonumber\\
 B(u)&=&g_\pi(u)-\varphi_\pi(u)\, ,
\end{eqnarray}
where
\begin{eqnarray}
h_{00}&=&v_{00}=-\frac{\eta_4}{3} \, ,\nonumber\\
a_{10}&=&\frac{21}{8}\eta_4 \omega_4-\frac{9}{20}a_2 \, ,\nonumber\\
v_{10}&=&\frac{21}{8}\eta_4 \omega_4 \, ,\nonumber\\
h_{01}&=&\frac{7}{4}\eta_4\omega_4-\frac{3}{20}a_2 \, ,\nonumber\\
h_{10}&=&\frac{7}{2}\eta_4\omega_4+\frac{3}{20}a_2 \, ,\nonumber\\
g_2&=&1+\frac{18}{7}a_2+60\eta_3+\frac{20}{3}\eta_4 \, ,\nonumber\\
g_4&=&-\frac{9}{28}a_2-6\eta_3\omega_3 \, ,
\end{eqnarray}
  $ C_2^{\frac{1}{2}}(\xi)$, $ C_4^{\frac{1}{2}}(\xi)$
, $ C_2^{\frac{3}{2}}(\xi)$ and $ C_4^{\frac{3}{2}}(\xi)$ are
Gegenbauer polynomials,
  $\eta_3=\frac{f_{3\pi}}{f_\pi}\frac{m_u+m_d}{m_\pi^2}$ and  $\rho^2={(m_u+m_d)^2\over m_\pi^2}$
 \cite{LCSR1,LCSR2,LCSR3,LCSR4,LCSR5,Belyaev94,PSLC1,PSLC2,PSLC3,PSLC4}.

\section*{Acknowledgments}
This  work is supported by National Natural Science Foundation,
Grant Number 10405009, 10775051, and Program for New Century
Excellent Talents in University, Grant Number NCET-07-0282, and Key
Program Foundation of NCEPU.

\end{document}